\documentclass{bioinfo}
\copyrightyear{2020} \pubyear{2020}

\access{Advance Access Publication Date: Day Month Year}
\appnotes{Manuscript Category}

\usepackage{xspace}
\newcommand{\code}[1]{\texttt{#1}\xspace}
\newcommand{\pkg}[1]{\textbf{#1}\xspace}
\newcommand{\proba}{\pkg{mlr3proba}}
\newcommand{\distr}{\pkg{distr6}}
\usepackage{xcolor}
\usepackage{listings}
\usepackage{color}
\usepackage{textcomp}

\begin{document}
\firstpage{1}

\subtitle{Subject Section}

\title[mlr3proba]{mlr3proba: An R Package for Machine Learning in Survival Analysis}
\author[Sonabend \textit{et~al}.]{Raphael Sonabend\,$^{\text{\sfb 1,}*}$, Franz J. Kir\'aly\,$^{\text{\sfb 1}}$, Andreas Bender\,$^{\text{\sfb 2}}$, Bernd Bischl\,$^{\text{\sfb 2}}$ and Michel Lang\,$^{\text{\sfb 2}}$}
\address{$^{\text{\sf 1}}$ Department of Statistical Science, University College London, London, WC1E 6BT, UK and \\
$^{\text{\sf 2}}$ Department of Statistics, LMU Munich, Munich, 80539, Germany.}

\corresp{$^\ast$To whom correspondence should be addressed.}

\history{Received on XXXXX; revised on XXXXX; accepted on XXXXX}

\editor{Associate Editor: XXXXXXX}

\abstract{\textbf{Motivation:} As machine learning has become increasingly popular over the last few decades, so too has the number of machine learning interfaces for implementing these models. Whilst many R libraries exist for machine learning, very few offer extended support for survival analysis. This is problematic considering its importance in fields like medicine, bioinformatics, economics, engineering, and more. \proba provides a comprehensive machine learning interface for survival analysis and connects with \pkg{mlr3}'s general model tuning and benchmarking facilities to provide a systematic infrastructure for survival modeling and evaluation.
\textbf{Availability:} \proba is available under an LGPL-3 license on \href{https://cran.r-project.org/}{CRAN} and at \href{https://github.com/mlr-org/mlr3proba}{https://github.com/mlr-org/mlr3proba}, with further documentation at \href{https://mlr3book.mlr-org.com/survival.html}{https://mlr3book.mlr-org.com/survival.html}.\\
\textbf{Contact:} \href{raphael.sonabend.15@ucl.ac.uk}{raphael.sonabend.15@ucl.ac.uk}}

\maketitle

\section{Introduction}

Survival analysis is the field of statistics concerned with the estimation of time-to-event distributions while accounting for censoring and truncation. \proba introduces survival modelling to the \pkg{mlr3} \citep{mlr3} ecosystem of machine learning packages. By utilising a probabilistic supervised learning \citep{skpro} framework \proba allows for multiple survival analysis predictions: predicting the time to an event, the probability of an event over time, and the relative risk of an event. \proba includes an extensive collection of classical and machine learning models and many specialised survival measures.

The R programming language \citep{Rstats} provides extensive support for both survival analysis and machine learning via its core functionality and through open-source add-on packages available from CRAN and Bioconductor. \proba leverages these packages by connecting a
multitude of machine learning models and measures for survival analysis. \proba currently supports simulation of survival data, classical survival models, prediction of survival distributions by machine learning, and support for high-dimensional data. Interfacing other packages in the \pkg{mlr3} family provides functionality for optimisation, tuning, benchmarking, and more.


\section{Implemented Functionality}
\label{sec:func}

A standard pipeline for survival analysis consists of: i) Defining a survival task as a set of features and survival outcome (time until the event and a censoring indicator); ii) Training a model on survival data, with the possibility of  optimisation via tuning of hyper-parameters; iii) Making predictions from the trained model on new data; iv) Evaluating the quality of predictions with survival-specific measures, possibly including visualisation.

\proba streamlines this process by: i) Standardising survival tasks, with the \code{Surv} object from the \pkg{survival} \citep{survival} package, into a single object capable of handling left-, interval-, and right-censoring (\code{TaskSurv}); ii) Unifying all survival learners (\code{LearnSurv*}) with iii) prediction objects that clearly distinguish model prediction types (\code{PredictionSurv}); iv) Unifying survival measures for different survival prediction types (\code{MeasureSurv*}).

Careful design and documentation of models and measures clearly demonstrate the predictions that can be made by models or evaluated by measures. Each model can predict one or more of: \code{response} - a survival time, \code{distr} - a survival distribution, \code{crank} - a relative risk ranking, and \code{lp} - a linear predictor. \code{distr} predictions are cast into standardized distribution objects using the \distr package \citep{distr6}, which allows clean post-processing, such as predicting survival and hazard functions, amongst other uses.

Any survival model implemented in \proba can be tuned via \pkg{mlr3tuning} \citep{mlr3tuning}, which includes several tuning methods (grid search, random search, generalized annealing and more) and termination criteria (based on iterations, runtime, and more) for nested resampling and optimisation on any survival measure. Additionally, all survival tasks and models can make use of \pkg{mlr3pipelines} \citep{mlr3pipelines} for pre-processing, such as feature selection and variable encoding, and post-processing, such as prediction compositions (see below). Full details for these methods are available in the mlr3book (\href{https://mlr3book.mlr-org.com}{https://mlr3book.mlr-org.com}).

\section{Implemented Classes}
\label{sec:obj}

\paragraph{Learners}
More than 20 survival learners are currently implemented. These range from classical statistical models to machine learning methods. For the former, the `usual' semi- and fully-parametric models are implemented, such as Cox PH \citep{cox_regression_1972} and AFT models, as well as more advanced flexible spline methods \citep{Royston2002} and penalized regression. Machine learning methods include random survival forests \citep{ishwaran_random_2008} (conditional inference, relative risk, log-rank splitting), gradient boosting machines (with multiple optimisation methods) \citep{Buhlmann2007}, Van Belle's support vector machines \citep{VanBelle2011}, and artificial neural networks \citep{Kvamme2019} (including DeepSurv, DeepHit, Cox-Time). Inclusion of Python-implemented survival neural networks via \pkg{survivalmodels} \citep{survivalmodels} allows efficient cross-platform comparison of models.

\paragraph{Measures}
For comparison of different models, 19 survival measures are implemented in \proba. These include quantitative calibration measures, such as van Houwelingen's $\beta$ \citep{VanHouwelingen2000}, and visual comparisons of average distribution prediction to Kaplan-Meier. Implemented discrimination metrics include several measures of both concordance (e.g. \citep{Harrell1982} and \citep{Uno2011}) and time-dependent AUCs \citep{Heagerty2000}. Scoring rules are also implemented including the log-loss, integrated log-loss, integrated Graf (or Brier) score \citep{Graf1999}, and the Schmid/absolute score \citep{Schmid2011}. Several of these are implemented directly in \proba with an \pkg{Rcpp} \citep{rcpp} implementation for fast and reliable performance.

\paragraph{Pipelines}
Pipelines provide a way to combine multiple pre- and post-processing steps into an object that can be treated as a learner. Such pipelines can include general and survival-specific components. One particular use case is the (re-)casting of one prediction type to another. There are several different possible predictions that could be made by a survival learner that are not directly comparable, e.g. a relative risk cannot be directly compared to a survival distribution. Therefore \proba extends the capabilities of any survival model by including pipelines that transform one prediction type to another. The \code{distrcompositor} pipeline transforms \code{lp} or \code{crank} predictions into \code{distr} predictions. Users have the option to specify the baseline distribution estimator (any learner implemented in \proba) and the model form (proportional hazards, accelerated failure time, or proportional odds). Another useful pipeline is the \code{crankcompositor}, this transforms a \code{distr} prediction into a \code{crank} and/or \code{response} prediction using some summary measure of the distribution, for example the mean or median. Obtaining a survival time prediction from a distribution is simply a case of wrapping the model in the \code{crankcompositor} pipeline. By combining these two pipelines, any model in \proba can make any prediction type. The \pkg{mlr3pipelines} functionality allows tuning of these and further pipelines to find the optimal parameters for these compositions.


\section{Related Work}

There are an increasing number of machine learning packages across programming languages, including \pkg{caret} \citep{caret}, \pkg{mlr} \citep{mlr}, \pkg{tidymodels} \citep{tidymodels}, and \pkg{scikit-learn} \citep{scikit-learn}. However, functionality for survival analysis has been mostly limited to `classical' statistical models with relatively few packages supporting a machine learning framework.
R ships with the package \pkg{survival}\citep{survival}, which supports left-, interval-,and right-censoring, competing risks, time-dependent models, stratification, and model evaluation. However, the package is limited to classical statistical models, with no support for machine learning and limited support for formal comparison or non-linear models. The Python equivalent to this package is \pkg{lifelines}\citep{lifelines}, which is again limited to a few classical models.
\pkg{pec}\citep{pec} implements no models itself but instead interfaces with many different survival packages to create survival probability predictions. The package's main focus is on model evaluation via prediction error curves (`pec's) with little support for model building/training and predicting.
\pkg{skpro}\citep{skpro} is a probabilistic supervised learning interface in Python. \pkg{skpro} extends the \pkg{scikit-learn} \citep{scikit-learn} interface to probabilistic models and appears to be the only package (in any language) dedicated to domain-agnostic probabilistic supervised learning.  The interface provides an infrastructure for machine learning based survival analysis with design choices influencing \proba, but \pkg{skpro} does not currently support survival models.
\pkg{pysurvival} \citep{pysurvival} is another Python package, which implements classical and machine learning survival analysis models. The package has the advantage of being able to natively leverage specific neural network survival models, which are almost exclusively implemented in Python. Whilst not directly interfacing the \pkg{scikit-learn} interface, the package introduces unified functions for model fitting, predicting, and evaluation.
\pkg{scikit-survival} \citep{Polsterl2020} builds directly on \pkg{scikit-learn} to implement a few survival models and measures in a machine learning framework. Unlike \pkg{pysurvival}, no neural networks are included, thus the two packages complement each other well.

\section{Future Developments}
As of now, the package is limited to the single-event, right-censored setting. This is largely a limitation of the current implementations of the underlying learners. Future developments will focus on extensions to: stratified models, time-varying effects, left-censoring/truncation, interval censoring, competing risks, and multi-state models. A recently proposed framework could be used to support most of these tasks without modification of the underlying learners \citep{bender_general_2020}. Some extensions, however, might require updates to the learners.
The near-future roadmap includes:
\begin{enumerate}
\item Expanding \code{TaskSurv} to accommodate the settings above.
\item Extending learners to handle (some of) the more complex settings.
\item Adding a learner-agnostic reduction pipeline for competing risks.
\end{enumerate}

\section{Example}

The example below demonstrates how to benchmark three survival models and make use of the distribution compositor. Line 1: Essential packages are loaded, \proba always requires \pkg{mlr3}. Line 2: Extra packages are loaded, \pkg{mlr3learners} is required for the xgboost learner and \pkg{mlr3pipelines} is required for the distribution composition. Lines 3-4: Kaplan-Meier and Cox PH learners are initialized with default parameters. Lines 5-7: The XGBoost learner, which doesn't provide predictions for the survival probability, is wrapped in the \code{distrcompositor} pipeline to transform its ranking prediction to a probabilistic prediction. Line 8: Learners are combined into a list for use in the benchmark function. Lines 9-11: A task is created from a subset of the \code{rats} dataset from \pkg{survival}, the outcome is specified with the `time' and `event' arguments. Line 12: A three-fold cross-validation resampling scheme is specified. Line 13: The infrastructure for the experiment is automatically determined by supplying the task(s), learners, and resampling method. Line 14: Learners are resampled according to the chosen scheme and benchmarked. Line 15: Predictions are aggregated over all folds and scored with the integrated log-loss to provide a final comparison.

\begin{verbatim}
> library(mlr3); library(mlr3proba)
> library(mlr3learners); library(mlr3pipelines)
> kaplan = lrn("surv.kaplan")
> cox = lrn("surv.coxph")
> xgb = ppl("distrcompositor",
+   learner = lrn("surv.xgboost"),
+   estimator = "kaplan", form = "ph")
> learners = list(cox, kaplan, xgb)
> task = TaskSurv$new(id = "rats", 
+   backend = survival::rats[,1:4],
+   time = "time", event = "status")
> resample = rsmp("cv", folds = 3)
> design = benchmark_grid(task, learners, resample)
> bm = benchmark(design)
> bm$aggregate(msr("surv.intlogloss"))
\end{verbatim}

\section*{Acknowledgements}
RS receives a PhD stipend from EPSRC (EP/R513143/1). AB and ML are funded by the German Federal Ministry of Education and Research (BMBF) under Grant No.~01IS18036A and by Deutsche Forschungsgemeinschaft (DFG) within the Collaborative Research Center SFB~876 ``Providing Information by Resource-Constrained Analysis''.


\bibliographystyle{natbib}
\bibliography{mlr3proba}

\end{document}